\documentclass[fleqn,10pt]{wlscirep}
\usepackage{hyperref}
\usepackage{float}
\usepackage[super]{nth}
\usepackage{multirow}
\usepackage{soul}
\title{The Evolution Of Centralisation on Cryptocurrency Platforms}

\author[1,2,3]{Carlo Campajola}
\author[2]{Raffaele Cristodaro}
\author[2]{Francesco Maria De Collibus}
\author[2]{Tao Yan}
\author[2,3]{Nicol\`o Vallarano}
\author[2,3,*]{Claudio J. Tessone}

\affil[1]{DLT Science Foundation, London, United Kingdom}
\affil[2]{University of Zurich, Blockchain \& Distributed Ledger Technologies, Department of Informatics, Z\"urich, Switzerland}
\affil[3]{UZH Blockchain Center, Z\"urich, Switzerland}
\affil[*]{claudio.tessone@uzh.ch}

\begin{abstract}
More than ten years ago the blockchain was acclaimed as the solution to overcome centralised trusted third parties for online payments. Through the years the crypto-movement changed and evolved, although decentralisation remained the core ideology and the necessary feature every new crypto-project should provide. In this paper we study the concept of centralisation in cryptocurrencies using a wide array of methodologies from the complex systems literature, on a comparative collection of blockchains, in order to define the many different levels a blockchain system may display (de-)centralisation and to question whether the present state of cryptocurrencies is, in a technological and economical sense, actually decentralised.
\end{abstract}
\begin{document}

\flushbottom
\maketitle
\thispagestyle{empty}

\section*{Introduction}

%One advantage of the cryptocurrencies, is that they are technically accessible to anyone \cite{scott2016can}, this is due to the fact that opening a wallet does not require any onboarding process or check on the person or entity willing to hold tokens. As opposed to “common” bank accounts, where individuals have to go through mandatory compliance onboarding processes that scrutinize if the person has been involved in criminal activities or is a political exposed person (PEP) \cite{basel2001customer}. This procedure is particularly important when the financial institutions execute payments instructed by its clients, as they are obliged to adhere to anti-money laundering laws and regulations with the purpose of avoiding to transfer capitals to offenders \cite{basel2013sound}. This is due to the fact that traditional wire transfers rely on financial institutions acting as intermediaries, which results to being obliged of having a money account with a financial institution for transferring money to a person or entity.

%In the other hand, the above does not apply to cryptocurrency platforms, cryptocurrency transfers do not rely on a third party in order to be executed, Bob can send Alice tokens without any obliged check on the transfer and the counterparties and without the intervention or clearance of any bank or lender \cite{devries2016analysis}.

In the wake of the 2008 Great Financial Crisis, the whitepaper ``Bitcoin: A Peer-to-Peer Electronic Cash System" \cite{nakamoto2008bitcoin} was published by the pseudonymous Satoshi Nakamoto. The paper began with a direct attack to the state of electronic payments on the internet: a critique on the central role of third parties to provide trust among users and the complete absence of non-reversible payments online. After describing the problem the paper proceeds to present Nakamoto’s solution, a relatively old \cite{chaum1979computer} and up to then fringe technology: the blockchain, a distributed and decentralised ledger where information is maintained consistent across the network by a peer-to-peer consensus protocol and secured via cryptography. Bitcoin was the first and thus far most successful platform of what have become commonly known as ``cryptocurrencies", digital assets that rely on cryptographic security to solve the double-spending problem, typically through a decentralised ledger rather than a centralised authoritative server.

Following Bitcoin's success, many more cryptocurrency projects started, each one with its own specific quirk yet characterised by the same common goal: to build a decentralised payment system, available to anyone without the need of reciprocal trust (or a third party providing it). For this reason the proponents of Bitcoin and other cryptocurrencies typically call it a ``truly democratic" form of money \cite{bollier2015democratic}, as opposed to the ``tyrannical money" issued by governments.

The main purpose of our study is to question whether the initial aspirations behind the creation of Bitcoin have been fulfilled. As pointed out by Taleb \cite{nicholas2021bitcoin}, Bitcoin and the likes suffer from design problems that lead to intrinsic fragilities, which in turn can compromise their utility. In our work we investigate these fragilities: we perform a large scale comparative analysis of multiple cryptocurrency platforms, sourcing the data directly from the respective blockchains, and adopt methods from network and complex systems science to define and investigate the concept of decentralisation in cryptocurrencies systems.

The blockchains we analyse are Bitcoin, Ethereum, Bitcoin Cash, Litecoin, Dogecoin, Monacoin and Feathercoin. We analyse data ranging from the systems inceptions until December 2020, adopting state of the art methodologies \cite{fischer2021complex} to transform the raw blockchain data into a meaningful dataset of economic transactions between ``entities". Entities are intended to be the closest possible representation of distinct economic agents on the blockchain: they can correspond to individuals, groups of individuals or businesses that operate in the system, coping with the pseudonymous nature of cryptocurrencies.

The entities on the blockchain interact through transactions, i.e. by sending tokens to one another to exchange value. This interaction mechanism is naturally suitable for a network representation, which has been widely adopted in the literature \cite{kondor2014rich}, where entities are mapped to nodes and directed links are established following the transactions flow. This provides a new perspective to study a cryptocurrency as an interconnected system of agents \cite{newman2006structure} of which we are able to quantitatively define and measure the centralisation. As an example, the emergence of a giant connected component \cite{newman2002assortative} in the transaction networks of all the cryptocurrencies under study signals how all the users participate to the same economic community. Another interesting point is the detection of a core component \cite{barucca2016disentangling} of the network, a subset of highly inter-linked nodes which keep the network together, taking a middle-men positioning in the networks. Moreover we are able to measure the wealth of entities and relate it to their position in the transaction networks. Similar analyses have been performed in the past for the ``cumulative" network, i.e. the network containing all transactions that happened up to a given point in history \cite{kondor2014rich}: we go beyond by splitting the time-frame in weekly time-windows, thus reducing the reciprocal impact of transactions spread far apart in time. It is indeed not particularly reasonable to consider in the same static network transactions which happened over 10 years apart. A similar approach in this respect has been recently taken for Bitcoin \cite{bovet2019evolving, makarov2021blockchain}, and we generalise their analysis to six more platforms, providing an unprecedented quantitative overview of the crypto market.

In the following section we describe the impact that the identification of entities has on the structure of the blockchain data and of transaction networks. We then focus on the networks and their global properties, such as the distribution of degrees and the identification of the network core, and then tackle the problem of mining power concentration and how it affects the distribution of wealths in the system. We conclude by discussing the results. For readability purposes the technical details about the data and methods are reported in the Materials and Methods section, following the discussion.

\section*{Results}
%One of the preliminary results obtained by comparing the centrality measures between the clustered blockchain networks and the non-clustered ones are the longest strongly connected component (LSCC), as expected, it is unequivocal the fact that the analysis outcomes strongly differ between the two networks, in the following evidence \nameref{Fig:btc_lscc}, the plot provides the size of the LSCC over time related to the Bitcoin transactions, the size is represented in percentage, and the available data start from the first block up to the 660964, created on the \nth{12} of December, 2020. 

\begin{figure}[t]
    \centering
    \includegraphics[width=.49\linewidth]{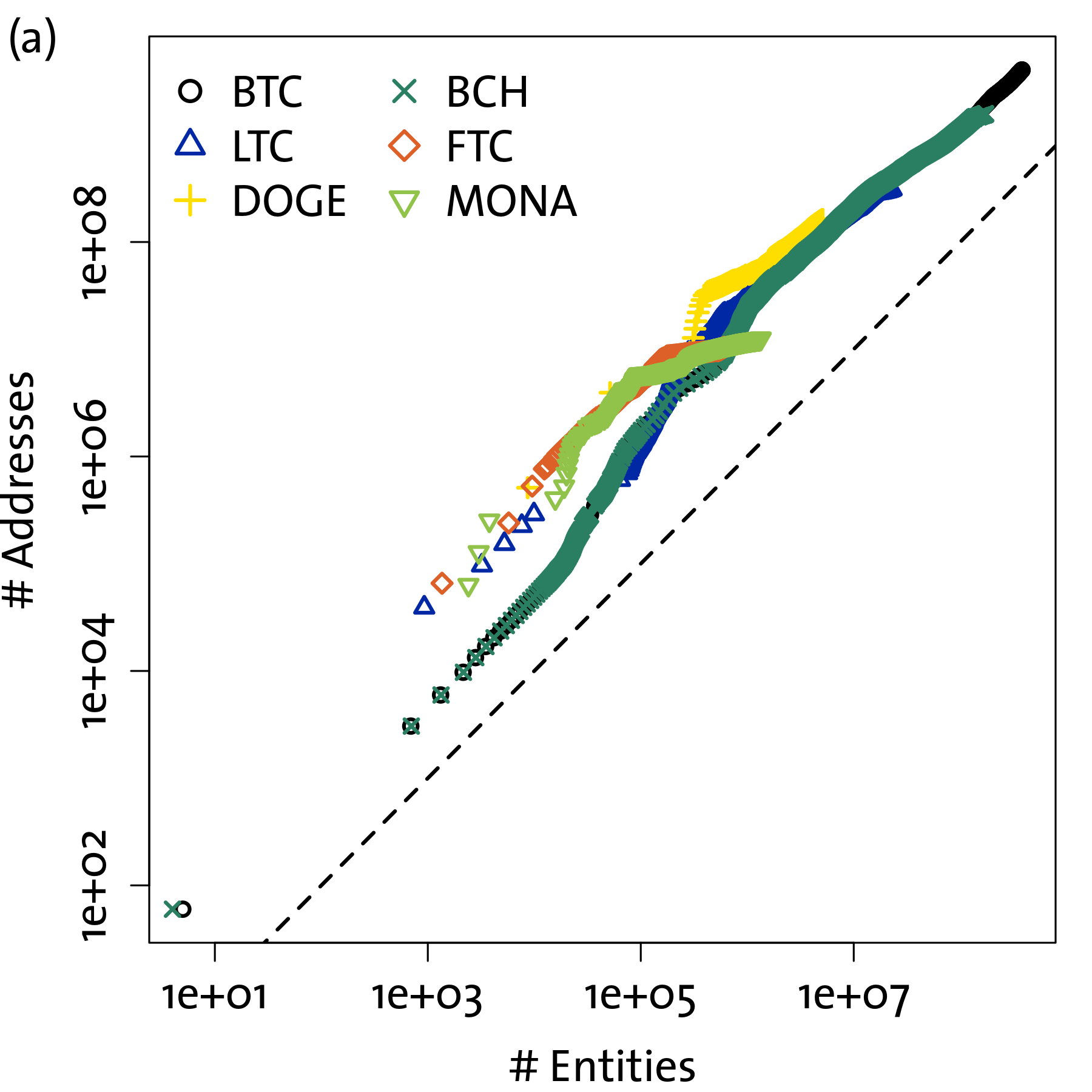}
    \includegraphics[width=.49\linewidth]{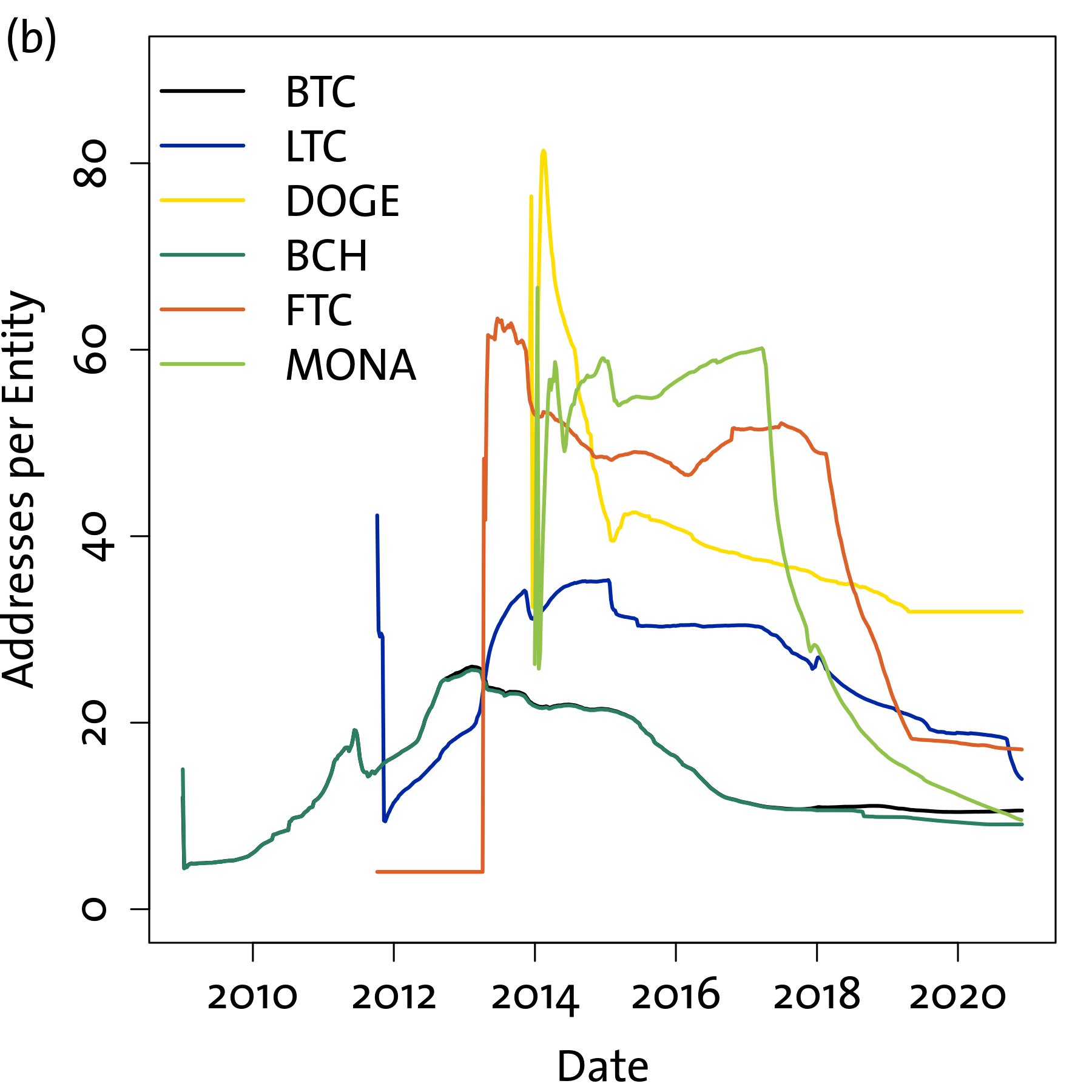}
    \caption{(a) Total number of addresses appeared on blockchain as function of total number of identified entities. Each plot point corresponds to the totals measured on the same block, with weekly frequency. Axes are logarithmic and the dashed line is reference for $y=x$. (b) Average number of addresses per entity over time.}
    \label{fig:nadd_nclust}
\end{figure}

\begin{figure}[t]
 \centering
 \includegraphics[width=.95\linewidth]{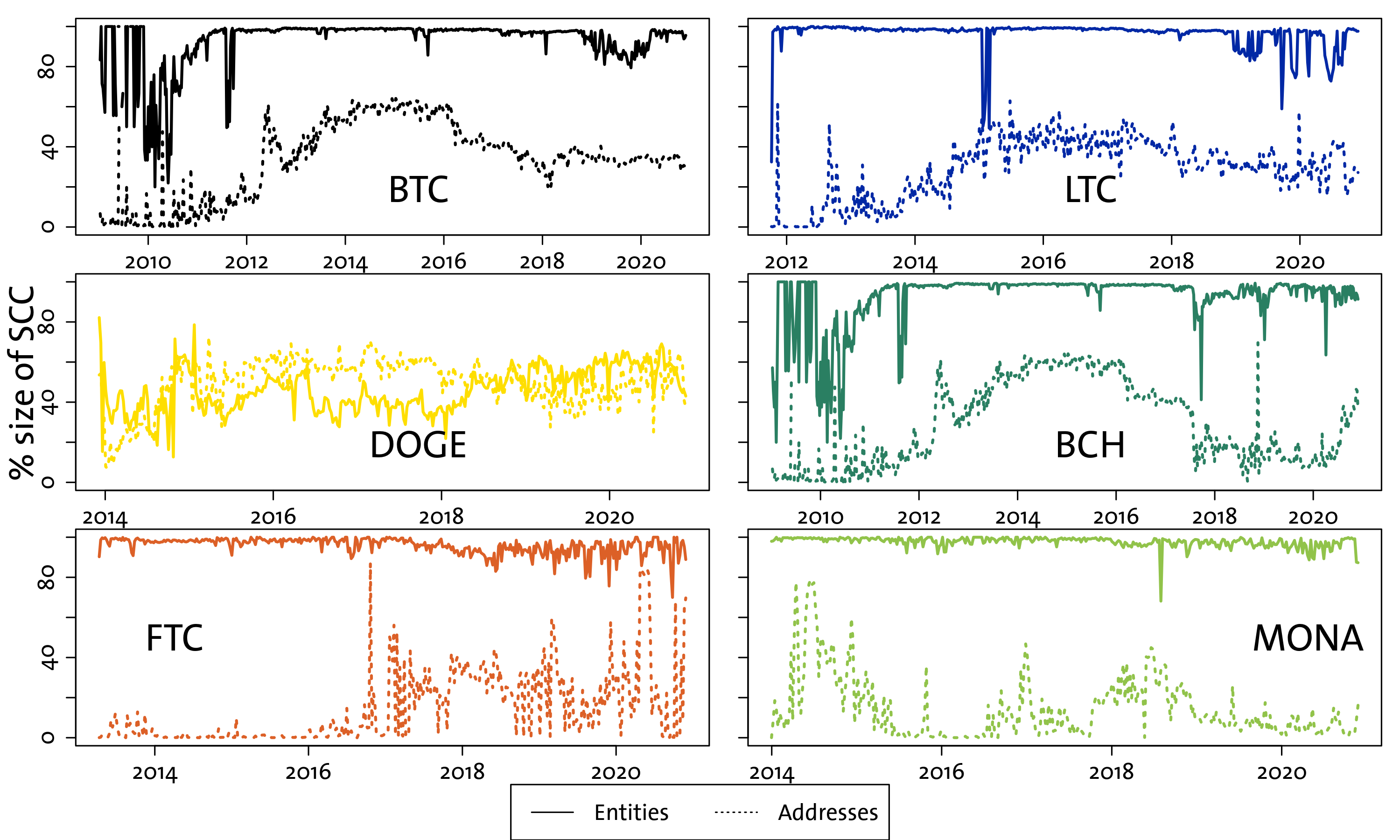}
 \caption{Size of the largest Strongly Connected Component in UTXO-based cryptocurrencies.}
 \label{fig:scc}
\end{figure}

%The results reported above provide graphically a rationale why the authors determined to apply a clustering algorithm to the blockchains examined in the paper, the study of the centralisation of the cryptocurrency platforms must take into consideration the data structure to encode transactions in UTXO based networks and the fact that active users may be willing to reduce their activities towards the network.

\paragraph{Clustering matters.} We begin by analysing the effect of applying address clustering on UTXO-based blockchain data. Figure \ref{fig:nadd_nclust}a shows the relation between the number of entities, identified by adopting the heuristic address clustering logic described in the Materials and Methods section, and the number of addresses that have appeared on the blockchain. Each data point refers to the totals observed up to a given block, with weekly frequency, and time ``flows'' from the bottom left to the top right of the panel. The results show strong similarities across the six platforms considered: this consistency is expected, as all these platforms share the same architecture in the address management protocols. New addresses are continuously generated with each transaction, and more active entities will accumulate more addresses on average. It is interesting to notice that there are several ``regimes'' in Figure \ref{fig:nadd_nclust}a, with the slope of the curve changing over time. We plot this slope, i.e. the average number of addresses per entity, as a function of time in Figure \ref{fig:nadd_nclust}b. Three regimes can be identified in the history of these cryptocurrencies: an early stage where the average grows, a relatively stable period between 2014 and 2017, followed by a decline after 2018. Interestingly, the transition from one regime to the other is marked by two of the well-known ``crypto bubbles", in 2013 and 2017, that were possibly the drivers of these shifts \cite{huber2022boom}. The first saw the creation of multiple so-called ``alt-coins", alternative cryptocurrencies running on their own blockchains which aimed to capitalise on the large influx of funds in the ecosystem, a class which includes all non-BTC assets in our analysis. In the period between the bubbles the average number of addresses owned by entities stayed at a relatively high value: relatively few new participants were entering the system since public interest had declined after the bubble burst, which means that transactions (and thus address creation) were primarily involving pre-existing entities. The picture changed after the 2017-18 bubble, when cryptocurrencies entered the mainstream of financial assets, receiving serious consideration by markets and institutions and leading to the consolidation of major service providers like leading exchanges Coinbase, Binance, etc. Since then public attention for the crypto ecosystem has stayed high and a large number of new entities entered the system, most of the times controlling just one or two addresses since they wouldn't use the blockchain to operate many transactions anymore. Indeed after 2018 blockchains started to experience congestion problems \cite{brown2021drives}, i.e. blocks reached capacity and thus adding new transactions to the blockchain became more costly. This created incentives to operate off-chain, mostly through service providers or so-called ``layer 2" technologies like the Lightning Network \cite{lin2021weighted}, thus reducing the average number of addresses per entity.

We can then consider the effect that address clustering has on the structure of transaction networks. We measure the size of the largest Strongly Connected Component (SCC) in the address networks and compare it to the one in the entity networks, shown in Fig. \ref{fig:scc} for the six cryptocurrencies of our analysis. Such a metric is essential to measure the fragmentation of the economy: if many nodes don't belong to the largest SCC it is a significant sign of decentralisation, as it means that there are no bridges across different communities; on the other hand if the vast majority of nodes belongs to the largest SCC it signals that such bridges exist and that a token can travel from one side of the network to the other. It is clear the address networks do not show the so-called ``giant component'' (i.e. a SCC that spans the whole network), which would suggest that these economies are highly fragmented and decentralised; however the illusion becomes apparent as soon as one looks at entity networks, where the largest SCC spans almost the entire network for most of the time. We should stress out the emergence of a Giant Connected Component is a necessary yet not sufficient condition for the economy to be centralised: a disconnected economy composed of isolated communities which keep money from flowing around cannot be centralised as a whole. 
A notable outlier in this is Dogecoin, where the largest SCC seems to be of about the same size in both representations, spanning approximately half the network. This observation then provides evidence that address clustering is a necessary step to take to get meaningful results about the economic structure of cryptocurrencies, as the obfuscation of identities through multiple addresses significantly affects the observations. In the following we will only present results on the entity networks for UTXO-based cryptocurrencies, unless specified otherwise.

\begin{table}[t]
\centering
\begin{tabular}{r|r|rrrrrrr}
  \hline
 & & BTC & LTC & DOGE & BCH & FTC & MONA & ETH \\ 
  \hline
\multirow{5}{*}{$\alpha$} & Minimum & 1.74 & 1.31 & 1.45 & 1.47 & 1.60 & 1.49 & 1.50 \\ 
  & Maximum & 10.99 & 5.84 & 3.83 & 9.89 & 4.81 & 7.69 & 3.42 \\ 
  & 1. Quartile & 2.42 & 2.43 & 2.45 & 2.28 & 2.33 & 2.46 & 1.68 \\ 
  & 3. Quartile & 2.93 & 3.08 & 3.09 & 2.94 & 3.18 & 3.21 & 1.81 \\ 
  %Mean & 2.71 & 2.77 & 2.78 & 2.65 & 2.74 & 2.85 & 1.82 \\ 
  & Median & 2.69 & 2.71 & 2.88 & 2.65 & 2.68 & 2.72 & 1.76 \\ 
   \hline
 $D_{Par}$ & Median & 0.03 & 0.05 & 0.05 & 0.04 & 0.08 & 0.10 & 0.03 \\ 
 $D_{Bin}$ & Median & 0.35 & 0.41 & 0.42 & 0.37 & 0.37 & 0.41 & 0.43 \\ 
\end{tabular}
\caption{Summary statistics on exponents of power-law fit of degree distribution on the weekly transaction networks.}\label{tab:degexp}
\end{table}

\begin{figure}[t]
    \centering
    \includegraphics[width=\textwidth]{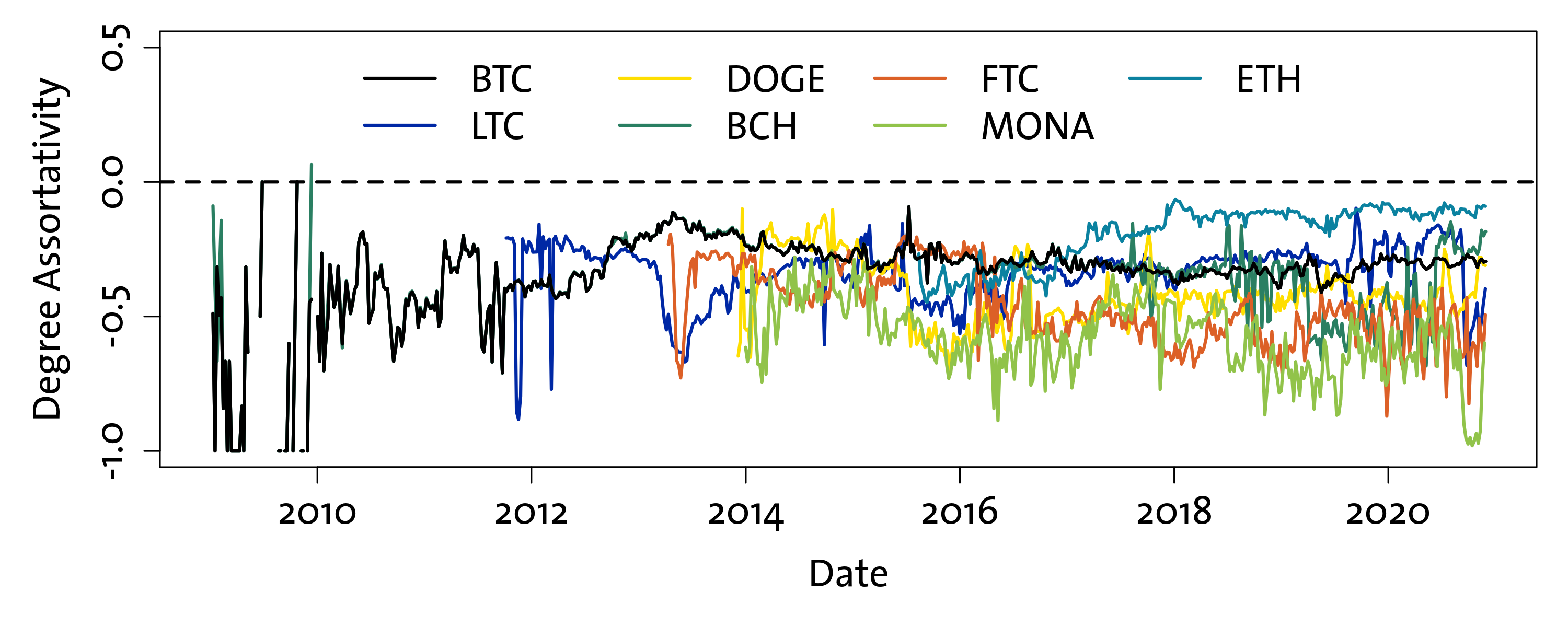}
    \caption{Assortativity coefficient for weekly transaction networks.}
    \label{fig:assor}
\end{figure}

\paragraph{Degree distributions.} The degree of a node represents the number of counterparts that entity exchanges tokens with, and is typically considered a measure of importance within the network. What is particularly relevant is to consider the distribution of degrees, as its shape is a direct consequence of the way in which the networks form. The degree distribution has been shown to be strongly skewed, akin to a power-law Pareto type distribution, for many traditional economic and financial systems such as interbank markets \cite{bargigli2015multiplex}, corporate payments \cite{sugiyama2005application} and in experiments \cite{tseng2010experimental}. This has also been observed in several analyses of cryptocurrency networks of transactions that represent the whole history \cite{kondor2014rich, de2021heterogeneous}, and our results broadly agree with the literature on the much shorter weekly timescale we select. We estimate the maximum likelihood parameters of a Pareto distribution and a binomial distribution for the degrees, which would be consistent with the null hypothesis of a Barabasi-Albert-type network or an Erdos-Renyi random network respectively, and find that the former is a much better description of the data than the latter in all of our samples. While we do not find that either distribution properly fits the data - indeed a Kolmogorov-Smirnov test rejects both null hypotheses at the $20\%$ level in the vast majority of weeks - we still find that the Kolmogorov-Smirnov distance is much lower for a Pareto-type distribution ($D_{Par}$ in Table \ref{tab:degexp}) rather than a binomial ($D_{Bin}$), hinting to the presence of fat tails. In Table \ref{tab:degexp} we report the statistics for the exponent $\alpha$ of a Pareto fit of the degree distribution, averaged over the whole sample. We find that in general the power-law exponent is between 2.5 and 3 for the UTXO-based cryptocurrencies - for which heuristic address clustering is available - and around 1.76 for Ether. This stark similarity across platforms points towards a scale-free structure of these transaction networks, with few hubs and large amounts of lowly connected nodes. This is further confirmed by computing the assortativity coefficient, which we report in Fig. \ref{fig:assor}. This is typically negative, with a tendency for smaller platforms like Monacoin and Feathercoin to show stronger disassortativity. This simple analysis is itself consistent with a strong centralisation in the flow of tokens, where the vast majority of transactions happens between low-degree nodes and large intermediaries like exchanges, custodians and service providers. Our observations largely agree with the evidence presented by \cite{makarov2021blockchain} on the Bitcoin transaction network, where the authors find that most of on-chain transactions (excluding self-transactions, which we also exclude) involve exchanges.

\begin{figure}[t]
    \centering
    \includegraphics[width=\textwidth]{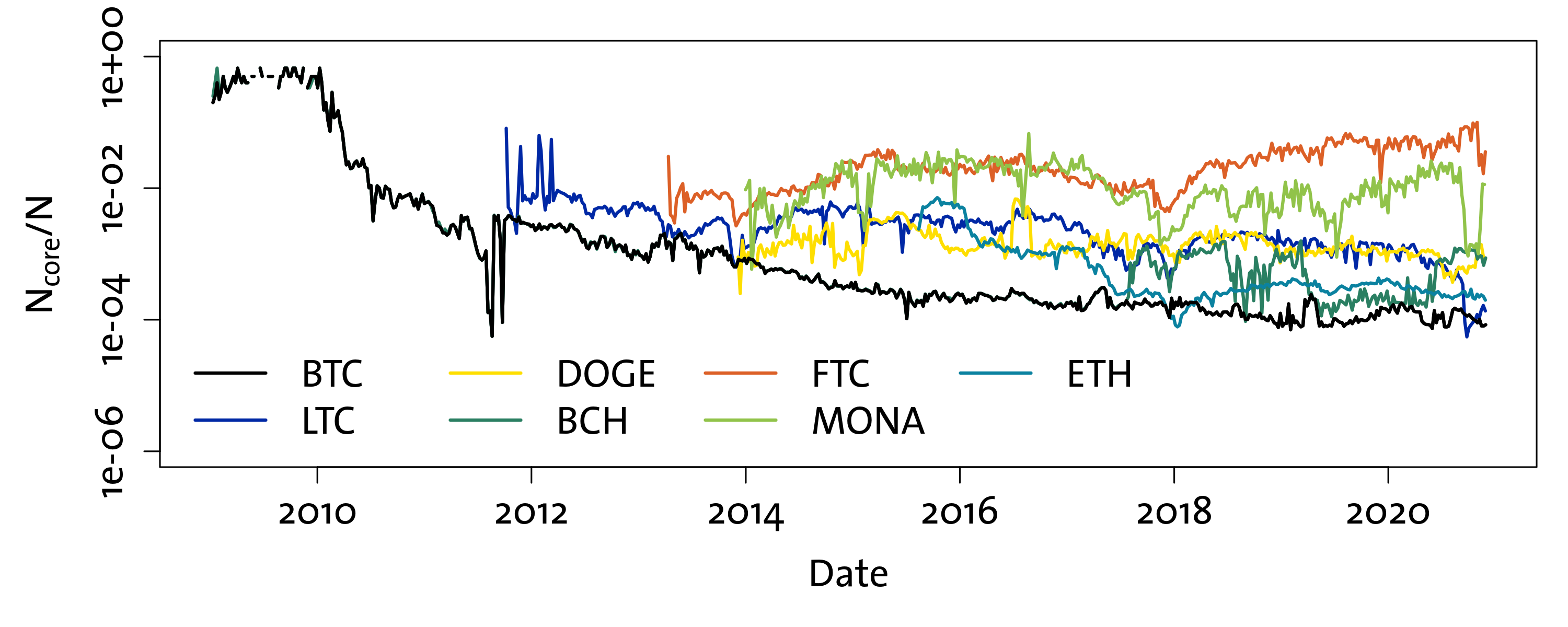}
    \caption{Fraction of nodes in the network core as a function of time.}
    \label{fig:coreperiph}
\end{figure}

\paragraph{Core-periphery structure.}
Economic and financial networks often exhibit the so-called ``core-periphery structure" \cite{borgatti2000models, barucca2016disentangling, bardoscia2017pathways}. This is a macroscopic property of the network, that presents a split between a minority of nodes (the ``core") with a strong connectivity between themselves and the remaining nodes of the network (the ``periphery") that are mostly connected to core nodes and have relatively few links to other peripheral nodes. We apply the core-periphery classification algorithm from \cite{lip2011fast} on the weekly transaction networks time-series and then consider the size of the core group as a fraction of the total size of the network. We report this quantity for each weekly transaction network in Figure \ref{fig:coreperiph}. We find that the relative size of the core has consistently decreased over time in all the analysed cryptocurrencies before 2018, and since then it has stabilised, excluding the particular cases of Monacoin and Feathercoin.
The fact that the relative size of the core of these networks has been shrinking is further evidence that the cryptocurrency economy has been shifting towards a centralised model, becoming much closer to the traditional financial system that it criticised heavily in its early days. A marked core-periphery structure points towards the transformation of the blockchain ledger into a kind of ``interbank market", where few large intermediaries move funds on behalf of their many clients much like bank transfers and brokerage happen in traditional banking. Indeed a shrinking core corresponds to a bigger periphery, i.e. the vast majority of entities (up to 99.99\% in the case of Bitcoin and Ethereum) is loosely connected to the rest of the network and relies on a small minority to intermediate transactions. This structure has been consolidating since 2018, consistent with our observations about the SCC and the amount of addresses per entity that we previously discussed.

The exception here, represented by Monacoin and Feathercoin, is mostly due to the drop in popularity these two platforms have experienced after the burst of the 2018 crypto bubble. Monacoin has also suffered from attacks on its mining protocol \cite{saad2019countering}, and the market capitalisation of both coins has not recovered to its 2018 peak. The size of the core of both cryptocurrencies is extremely small (less than 5 nodes from 2018 onwards) but the total size of the network is also relatively small, with few hundreds of active entities, thus leading to the observed results.

\paragraph{Mining concentration.} An analysis of centralisation on cryptocurrency platforms would not be complete without reporting measures of mining power concentration. All blockchains in this study are based on the Proof-of-Work (PoW) consensus mechanism, where the right to add a new block to the blockchain is granted to whoever finds the solution to a cryptographic puzzle. These entities are typically called ``miners", and have an incentive to operate given by a \textit{coinbase transaction} that mints new tokens attributed directly to the miner. The consistency of blockchain data relies on the assumption that mining power is distributed and decentralised: if at any point in time a single miner (or a pool of miners) holds more than $50\%$ of the total computational power devoted to solving the PoW puzzle, the blockchain can be subjected to a so-called ``51\% attack" where the majority miner writes false information on the blocks (e.g. double spending transactions) without anybody being able to counteract, since the majority miner will always produce the longest chain with no opposition. A short-term ``51\% attack`` may lead to a huge decrease of confidence in a blockchain's content; the persistence of such an attack could mean the end of a blockchain project.

\begin{figure}[t]
    \centering
    \includegraphics[width=\textwidth]{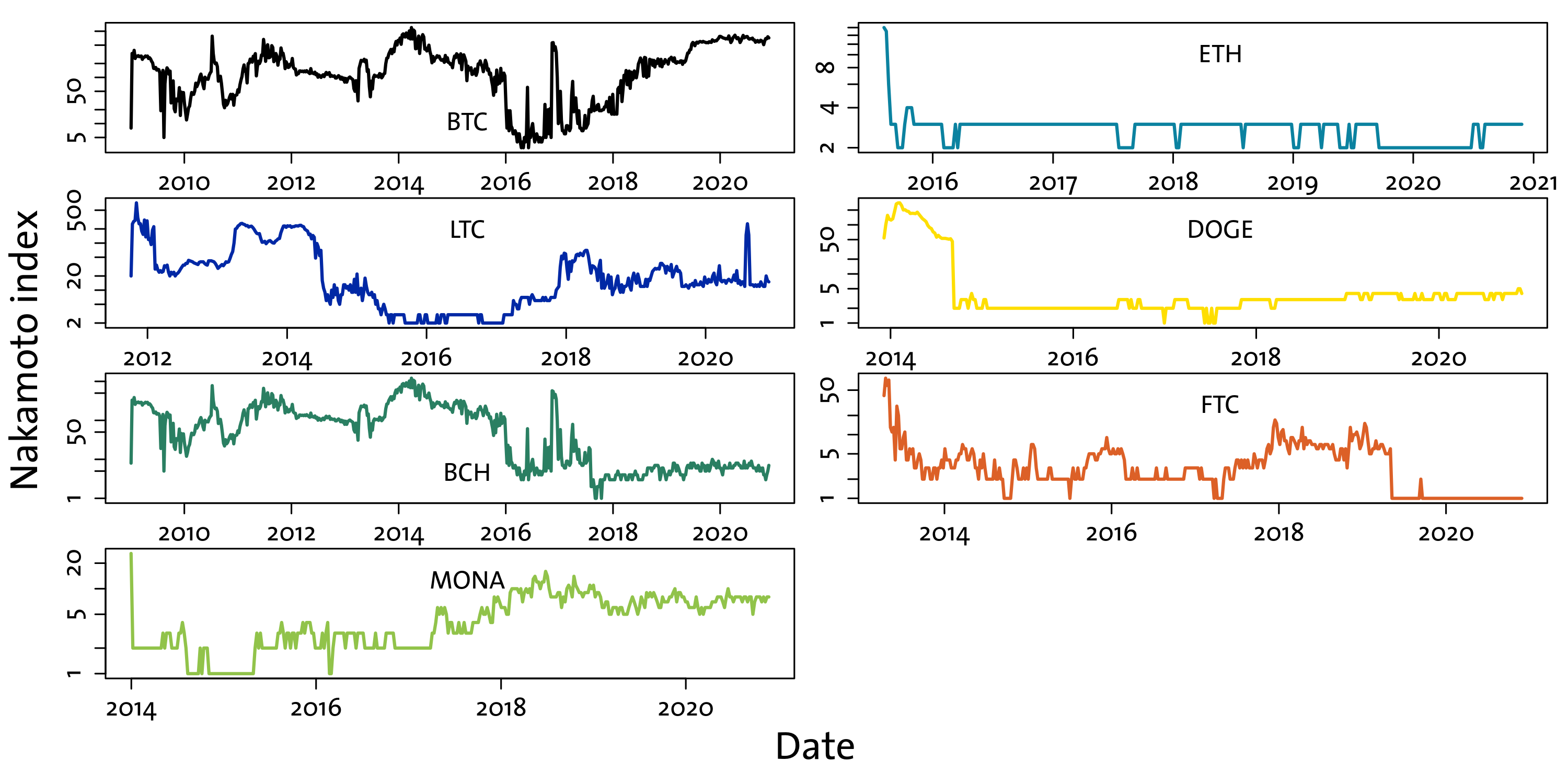}
    \caption{Nakamoto index estimated weekly for the analysed cryptocurrencies.}
    \label{fig:naka}
\end{figure}

The Nakamoto index provides us with a measure of the system's distance from a 51\% attack: named after the pseudonymous author of the Bitcoin whitepaper, it is defined as the minimum number of miners that need to coordinate at any point in time to run a $51\%$ attack. There have been several reports of hashing power centralisation on Bitcoin \cite{alsabah2020pitfalls, makarov2021blockchain}: based on blockchain data alone, we are able to estimate this index for all the analysed cryptocurrencies, and we show the result in Figure \ref{fig:naka}. We estimate the hashing power of miners by measuring the fraction of blocks they mine in a week, and then calculate the Nakamoto index over that same week. Miners are identified as the entities receiving the coinbase transaction: differently from other parts of this work, in Figure \ref{fig:naka} we show results that are obtained using only the multiple inputs heuristic method for address clustering, in order to avoid merging too many pools and miners together. We report similar results for the combined heuristics in the Supplementary Information.

The majority of analysed platforms shows worrying results. The most striking case is Ethereum, with a Nakamoto index that is almost never above 3 and in some weeks drops to 2: this is particularly concerning since Ethereum is used as the base protocol for many decentralised applications, including decentralised exchanges and other decentralised finance instruments which could introduce additional incentives for miners to coordinate in a $51\%$ attack \cite{daian2020flash, piet2022extracting}. The picture looks marginally better for Bitcoin and Litecoin, where the mining power appears relatively less centralised (although it still is very concentrated at times), while on Dogecoin, Feathercoin, Monacoin and Bitcoin Cash (after it forked from Bitcoin in August 2017) the Nakamoto index is always low, and at times even goes to 1, meaning a single miner could have (and maybe has) successfully run a $51\%$ attack on some weeks. Our evidence qualitatively agrees with previously presented models \cite{alsabah2020pitfalls} and evidence for Bitcoin \cite{makarov2021blockchain}, where in the latter the authors had additional information available from deanonymisation services specific to Bitcoin, which we don't have access to.

\begin{figure}[t]
    \centering
    \includegraphics[width=.405\textwidth]{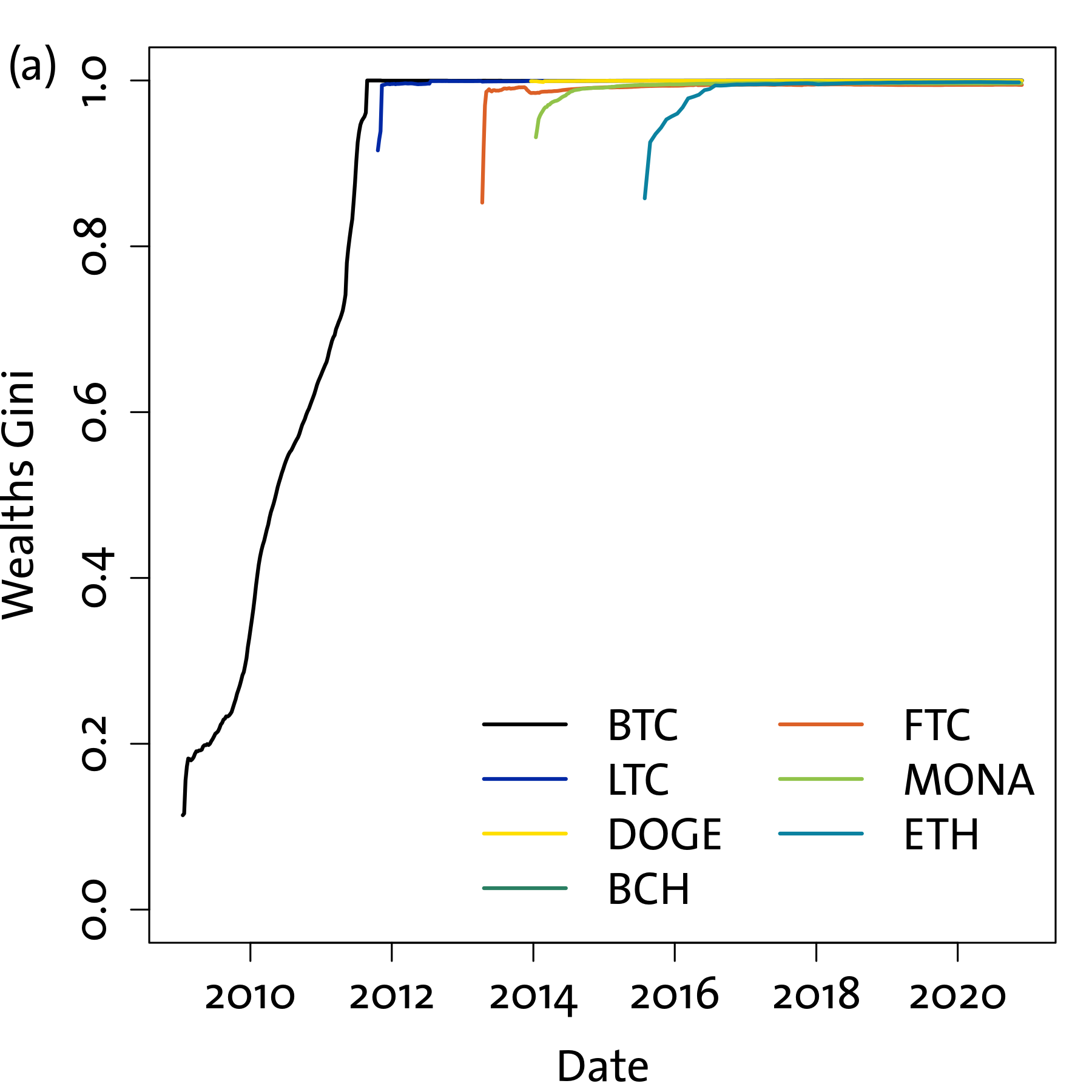}
    \includegraphics[width=.53\textwidth]{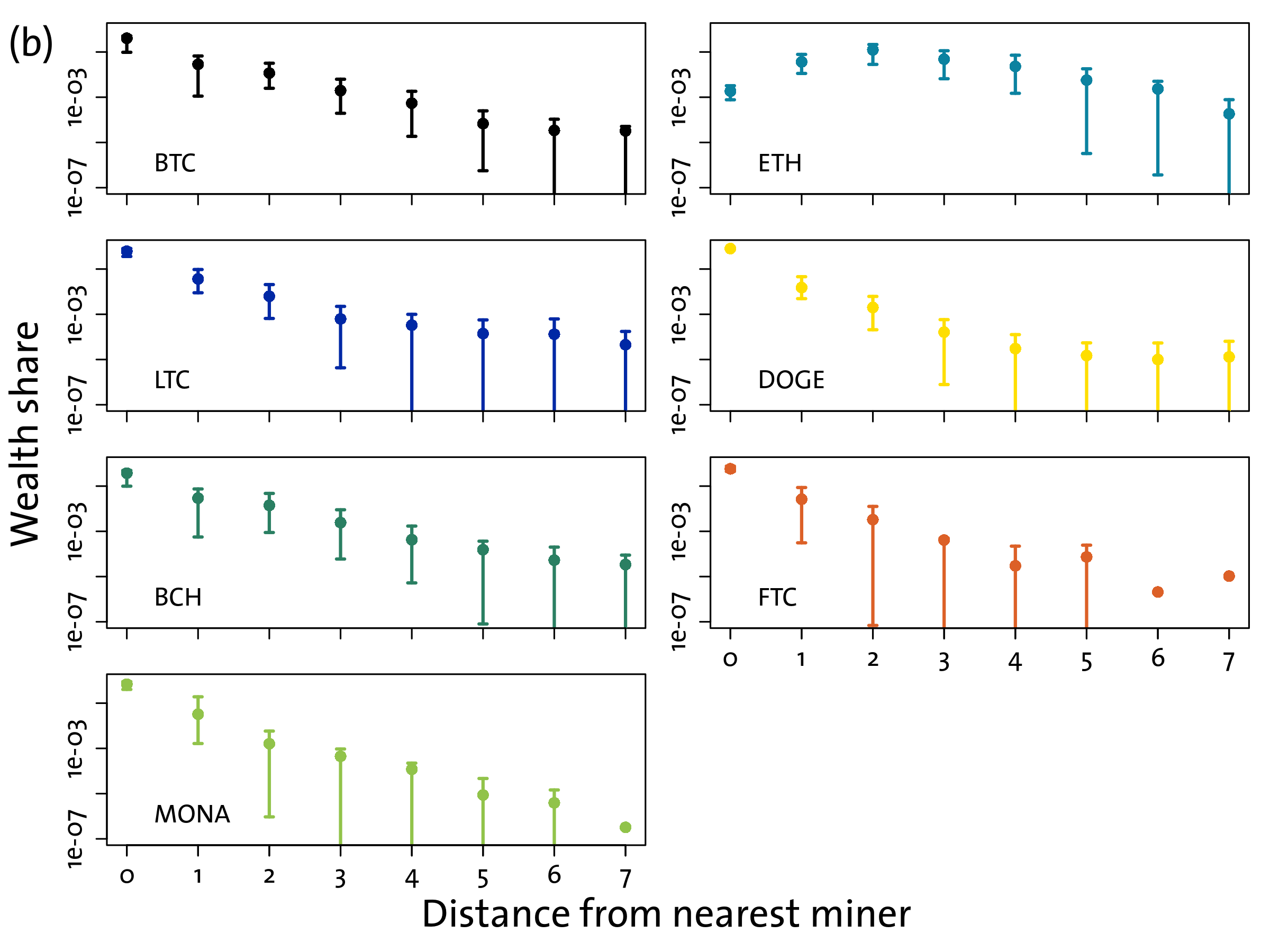}
    \caption{Wealth inequality and spatial distribution. (a) Evolution of the Gini index of wealths in analysed cryptocurrencies. (b) Distribution of tokens as a function of distance from the miner. Plot points identify the mean fraction of wealth, averaged over time, that is held by nodes at a given distance from the nearest miner. Error bars give the 90\% confidence band, i.e. in 90\% of weeks the fraction of tokens held at a given distance is within the error bars.}
    \label{fig:wealthdist}
\end{figure}

\paragraph{Wealth inequality and spatial distribution.} Our final analysis concerns the distribution of tokens among entities. We define as wealth the balance held by an entity (address or cluster of addresses) at a given point in time, representing the funds that they are entitled to spend according to the blockchain ledger. To avoid considering inactive wallets, our statistics only consider addresses or entities with non-zero balance. Clearly, if cryptocurrencies were used simply as a peer-to-peer payment system as implied by the Bitcoin whitepaper \cite{nakamoto2008bitcoin} (and many other whitepapers), the level of centralisation and inequality of wealths should not be very different from what is measured in traditional economic systems \cite{davies2018comparing}. This is very far from what can be measured on the system we analyse. In Figure \ref{fig:wealthdist}a we show the Gini coefficient of wealths, measured weekly on each cryptocurrency throughout our sample. This coefficient is a standard measure of economic inequality and is related to the area between the Lorenz curve - which plots the cumulative share of wealth as a function of the fraction of population owning that share - and the equality line - i.e. the straight 45 degrees Lorenz curve. A Gini coefficient of 0 means perfect equality, i.e. each individual has the same wealth, whereas a Gini coefficient of 1 means total inequality, with one individual owning everything. Remarkably, only Bitcoin in its early years has a Gini coefficient that is lower than the world average wealth inequality Gini coefficient (which is around 0.9, according to \cite{davies2018comparing}). After 2012, for all the cryptocurrencies in our analysis the Gini index is basically indistinguishable from 1, meaning that the distribution of wealth is extremely unequal across active wallets. We want to stress that this is not due to having many zero-wealth entities in the sample, as we remove these before calculating the Gini coefficient.

Finally, we characterise the spatial distribution of wealth on our transaction networks. To this end we calculate, for each weekly transaction network, the minimum distance that each node has from a miner. Miners are special entities in cryptocurrencies because they are the ones that generate new tokens, as every time they successfully solve the Proof of Work validation puzzle they are entitled to a reward that is created by the protocol. This means that all tokens start their ``life" in a miner's wallet and then diffuse from there into the rest of the economy, which is why they are a meaningful reference node to consider in transaction networks. %We identify active miners as the entities that received the reward from the coinbase transaction of blocks that are mined before the week on which the network is built. 
We then proceed to calculate the amount of wealth that is held by nodes at the same distance from a miner, normalised by the total amount of cryptocurrency circulating, and then consider this quantity across all weeks in our sample. We plot the result in Figure \ref{fig:wealthdist}b. The plot points represent the average wealth share owned by entities at a given distance from a miner, while the error bars show the 90\% confidence band, i.e. the range of values within which the wealth share lies in 90\% of weeks in our sample. We find remarkable similarities between the analysed cryptocurrencies: the average wealth share is a decreasing function of distance from miners, with a behaviour that is close to an exponential, and miners (at distance 0) are almost always the wealthiest nodes. The only platform where this doesn't happen is Ethereum, where the largest wealth share is at a distance of 2 from miners. This is likely due to the difference in the accounting protocol between UTXO-based systems and Ethereum's account-based system, but we lack conclusive evidence about this. What appears clear from our analysis though is that the largest fraction of wealth is held by relatively few entities - as summarised by the Gini coefficient - that are in the immediate vicinity of miners, with a tail that is quickly vanishing as distance grows.

\section*{Discussion}

In this paper we have presented multiple evidences of the centralised nature of major cryptocurrencies, despite their statutory technological decentralisation. We described how the protocol of Bitcoin and similar UTXO-based blockchains works to preserve anonymity and how this can be partially reverse-engineered to cluster addresses that are likely belonging to the same entity, and how switching from a raw address-based description to an entity-based description radically changes the observations. In particular we found interesting dynamics in the evolution of the average number of addresses per entity, which is remarkably similar across platforms and experiences breakpoints in conjunction with large price movements. A first circumstantial evidence of the centralisation process comes from these dynamics, namely the decrease of the average number of addresses per entity after 2018, together with the consolidation of cryptocurrencies as popular speculative assets and the growth of large intermediaries. The result is consistent with a shift in the platforms usage, with new entities entering the system through centralised access points rather than independently.

The introduction of weekly transaction networks allowed us to analyse the structural properties of cryptocurrency economies. We found further evidence of the importance of address clustering by measuring the size of the largest Strongly Connected Component (LSCC) of the networks: while on address networks the LSCC contains less than half of active addresses, when considering entities the LSCC includes almost all the nodes of the networks. We take this as a further sign that there are entities which control multiple addresses, acting as ``bridges" between different sub-components of the economy. The networks we studied exhibit a fat-tailed degree distribution, with few entities participating in a large number of transactions each week while the vast majority only performs very few, and a negative assortativity coefficient, which constitutes further evidence of the existence of large intermediaries that centralise the flow of tokens \cite{campajola2022microvelocity}. Furthermore these networks exhibit a marked core-periphery structure, much like other trading networks from traditional economic and financial systems, and as time goes on the cores become increasingly smaller compared to the total size of the network.

Finally we focused our analysis on the role of miners, which are particularly important actors in these systems as they hold the responsibility of keeping the ledger consistent. We find that in all the studied platforms the mining power, namely the estimated share of hashing rate that each miner holds, is worryingly concentrated in the hands of few entities to the point that the Nakamoto index often drops to few units, implying a high risk of a successful $51\%$ attack on the blockchain. We also found that miners are typically the wealthiest entities in the system, and that the stark inequality in wealths is incompatible with the typical narrative of a decentralised, peer-to-peer payments system.

Overall we argue that these economies have grown into something that is very different from what they were originally designed to be. The popular narratives, also seen in some mainstream media, that cryptocurrencies are a decentralised solution for payments where financial institutions hold no power over money creation and distribution, are largely false. Similar processes have been reported in second-layer solutions, such as the Lightning Network \cite{Lin2020,lin2021weighted}. As a matter of fact, we see that miners and intermediaries hold a huge power in these systems, both technically and economically speaking, which turns them into an unregulated financial sector, a new ``shadow banking" system \cite{aramonte2021defi}. This unchecked concentration of power is potentially dangerous for the many retail investors that have holdings in cryptocurrency, and proper regulation regarding the activities and products that service providers can offer is overdue.

\section*{Materials and Methods}

\begin{table}[t]
    \centering
    \resizebox{\textwidth}{!}{\begin{tabular}{r|c|c|c|c|c|c|c|}
    \hline
        Blockchain & BTC & ETH & BCH & LTC & DOGE & MONA & FTC \\
        Height & 661386 & 12950000 & 664161 & 1969743 & 3927743 & 2212711 & 3491853 \\
        Date of collection & 14.12.2020 & 03.08.2021 & 02.12.2020 & 23.12.2020 & 07.10.2021 & 07.01.2021 & 07.01.2021 \\ \hline
    \end{tabular}}
    \caption{Summary information on data collection from public blockchains.}
    \label{tab:data}
\end{table}

\paragraph{Data.}
Blockchain data for the seven cryptocurrencies of our analysis is sourced directly from the respective networks by running consensus nodes, collecting data at block heights reported in Table \ref{tab:data}. The figures presented in the paper show results until the last date that is available for all blockchains, December 2, 2020.

The blockchain protocols analysed in this study belong to two different families when it comes to their accounting standard: the Unspent Transaction Output (UTXO) standard and the account-based standard, the latter being specific of Ethereum alone in our sample, while all others belong to the former. In UTXO blockchains transactions are recorded as transfers of \textit{access rights} to a certain amount of tokens, stored on an address: this is done by referencing one or more addresses currently owned by the transaction sender, and directing the related tokens to one or more addresses owned by the transaction receiver. Typically the first set of addresses is called the ``input" of the transaction, while the second is its ``output". Outputs can only be used (``spent") once as inputs to a transaction, hence the name of Unspent Transaction Output accounting.
Ownership of addresses is proven by public-private key cryptography: in order to access funds from an address, the private key is needed to sign the transaction in a way that any other user can verify via the public key. The same entity can own any number of addresses, thus making tracing of transactions between entities more challenging, as there is no guarantee that a transaction involves addresses from one, two or many different entities. Multiple heuristic algorithms have been developed over the years to exploit common patterns in transactions input-output structures to cluster together addresses that are likely to belong to the same entity. We adopt multiple combinations of these algorithms in our analysis, and report the details below.

In contrast to the UTXO model described above, the account-based blockchains use a different and perhaps more intuitive mechanism for exchanging amounts of cryptocurrencies and other cryptoassets. Instead of holding the private key that can unlock the unspent output for a new transaction where it will be used as input, account-based blockchains have their public key directly transformed as an address, and that address has a specific balance associated. There is then a reduced need to apply address clustering to account-based blockchains, since new addresses are not often created within the same wallet.
The first and up to the time of writing most successful example of account-based blockchains is Ethereum, with its native cryptocurrency Ether \cite{ether}. %In contrast to Bitcoin, Ethereum is meant to be Turing Complete, hence its smart contracts (i.e. programs) can absolve any possible purpose, taking in consideration the limits of programming for blockchain, a very specific application of distributed computing \cite{mastering-ethereum}.

%Our analysis is conducted on 6 blockchains which are Bitcoin, Bitcoin Cash, Litecoin, Dogecoin, Monacoin and Feathercoin. We collected the transaction data from the beginning of the blockchain to December 2020. To get the data we firstly installed the node client for each blockchain, Amount(table(time, number of transactions,amount)), then we used blocksci to download the transaction data. The transactions data includes timestamp, from\_address, to\_address, values.

%\begin{figure}[!htbp]
%\centering
%\includegraphics[scale = 0.21]{utxo.png}
%\caption{UTXO based transaction}
%\label{fig:utxo}
%\end{figure}

\paragraph{Address clustering in UTXO blockchains.}
The entities, i.e. clusters of addresses, have been generated using the C++ library BlockSci, which is an open-source software package for blockchain analysis \cite{blocksci}. We combined multiple heuristic algorithms that are provided within the library to produce our own heuristic methods; we briefly summarise them below and direct the reader to BlockSci's documentation\footnote{https://citp.github.io/BlockSci/} for further details. Most of the methods rely on the identification of \textit{change addresses}, i.e. outputs of a transaction that belong to the sender, to which excess input is directed. Indeed the protocol design imposes that the input value of a transaction exactly matches the output value plus the fee, which makes change addresses very common.

The heuristic methods we adopted are the following:
%The methodologies applied to the various blockchain platform transactions in order to compose the users transacting networks are two, which differs slightly, essentially the first methodology takes into consideration three  different static analysis on the transactions, whereas the second methodology applied the same three heuristics as the previous one plus an additional dynamic analysis. 

\begin{itemize}
    \item Multiple inputs - if multiple addresses appear as inputs to the same transaction, they are likely to belong to the same entity that had to sign them all to create the transaction;
    \item Reuse of an address - if an address appears as an input but also as an output, the address has been reused as change;
    \item Optimal change: if the transaction has multiple inputs and the value of exactly one output is lower than any of the inputs, then it is likely the change. This is based on the argument that if a bigger output was the change, then one or more inputs would have not been needed to transfer the same amount to another wallet;
    \item Creation of a new address - in case an output address is appearing for the first time on the blockchain, that address has likely been generated to be the change address;
    \item Peeling chains - a transaction is considered a potential peeling chain if it includes one input and two outputs, and the chain is identified if one of the outputs is used as input to a future transaction with the same structure. In that case, the output that continues the chain is the change address. These chains are commonly seen when addresses with large UTXO values are spent in many smaller value transactions.
            
\end{itemize}

%In the first methodology, the cluster is determined in case the heuristics optimal change and creation of a new address agree on a unique change output, additionally, this result should not diverge from the heuristic established from the reuse of an address result.
%Below a table providing an example on how the outcomes of the heuristics have been managed, the resulting addresses have been clustered with the input addresses of the transaction, implicating that they belongs to the same user creating the transaction itself.  
%\begin{table}[!h]
%	\centering
%	\begin{tabular}{|c|c|c|c||c|}
%		\hline
%		& \textbf{Reuse Address} & \textbf{Optimal Change} & \textbf{Creation Address} & \textbf{Clustering Address} \\ [0.5ex] 
%		\hline\hline
%		1. & Address01 & NoResult & NoResult & Address01\\ [1ex] 
%		\hline
%		2. & Address01 & Address02 & Address03 & NoResult\\ [1ex] 
%		\hline
%		3. & Address01 & Address02 & Address01 & NoResult\\ [1ex]
%		\hline
%		4. & NoResult & Address01 & Address01 & Address01\\ [1ex] 
%		\hline
%		5. & NoResult & Address01 & Address02 & NoResult\\ [1ex] 
%		\hline
%	\end{tabular}
%	\caption{\small \centering Possible results for methodology 1}
%	\label{tab:heurMeth1}
%\end{table}

\begin{figure}[t]
    \centering
    \includegraphics[width=\textwidth]{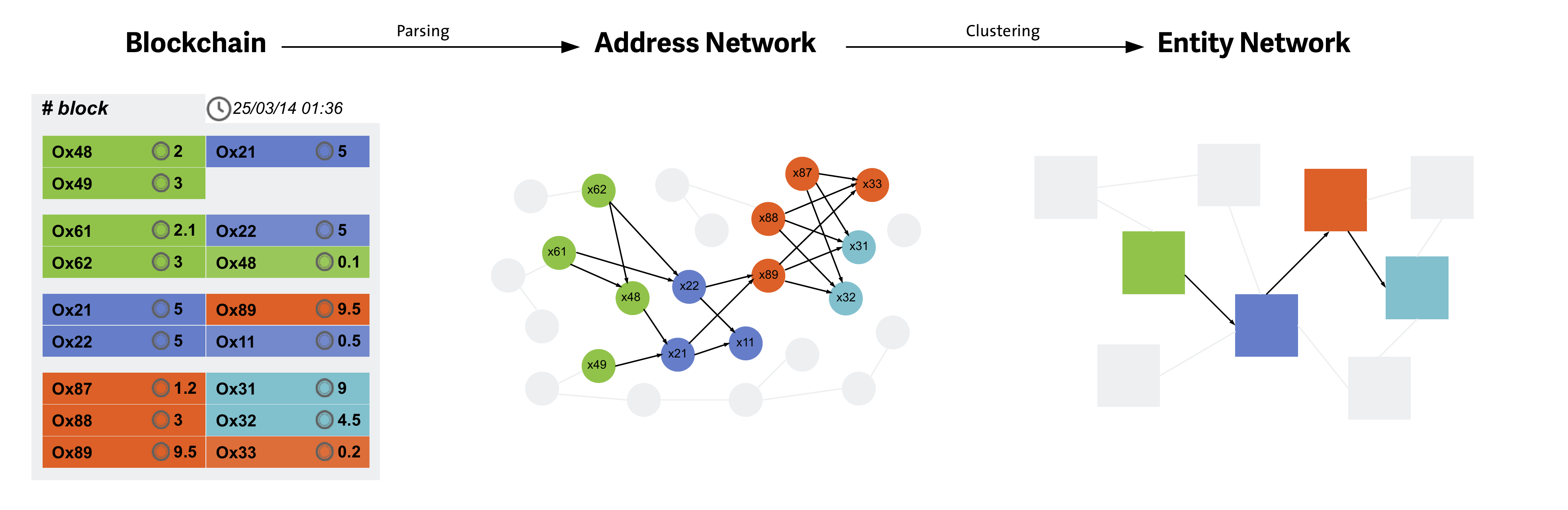}
    \caption{Schematic representation of transaction networks construction. On the left, an archetypal UTXO blockchain block is represented, with coloured transaction input/outputs identifying clusters according to common heuristics. This is parsed through BlockSci into the middle representation, the transaction network at the addresses level, which is then further reduced to the entity network on the right when multiple addresses/nodes belonging to the same entity are merged.}
    \label{fig:clust}
\end{figure}

Our main clustering methodology combines the above heuristics through logical operators that aim to reduce the chance that a false positive occurs. In particular we identify a change address only if all heuristics identify the same address as potential change, except for peeling chains which can never be compatible with other patterns by design. Hence a change address is identified if it is the only one that respects one of the following conditions: i) it is a reused input address; ii) it is a new address with a value smaller than any of the inputs; iii) it is the continuation of a peeling chain. If a change is identified, it is added to the same cluster to which the input addresses belong. A robustness analysis adopting a more conservative choice of clustering strategy is presented in the Supplementary Information.

%\begin{table}[H]
%	\centering
%	\begin{tabular}{|c|c|c|c|c||c|}
%		\hline
%		& \textbf{Reuse Address} & \textbf{Optimal Change} & \textbf{Creation Address} & \textbf{Peeling Chain} & \textbf{Clustering Address} \\ [0.4ex] 
%		\hline\hline
%		1. & Address01 & Address02 & Address03 & NoResult & Address01 \\ [0.9ex] 
%		\hline
%		2. & Address01 & Address02 & Address02 & Address03 & NoResult\\ [0.9ex] 
%		\hline
%		3. & Address01 & Address02 & Address02 & Address01 & NoResult\\ [0.9ex]
%		\hline
%		4. & NoResult & Address01 & Address01 & NoResult & Address01\\ [0.9ex] 
%		\hline
%		5. & NoResult & Address01 & Address02 & Address03 & Address03\\ [0.9ex] 
%		\hline
%		6. & NoResult & Address01 & Address01 & Address02 & NoResult\\ [0.9ex] 
%		\hline
%	\end{tabular}
%	\caption{\small \centering Possible results for methodology 2}
%	\label{tab:heurMeth2}
%\end{table}

%\subsection{User Network and centrality measures}

\paragraph{Transaction networks.}
We construct transaction networks to obtain a general representation of different blockchains. We choose a weekly aggregation to remove intraweek seasonalities, as we notice that the number of active entities and transactions is lower during weekends. We then consider all transactions between entities - i.e. removing self-transactions like change outputs - happening in a given week, adding the involved entities as nodes and connecting them through directed links from inputs to outputs. This process is summarised by Figure \ref{fig:clust}. Differently from previous literature on cryptocurrency transaction networks \cite{kondor2014rich, de2021heterogeneous}, we consider non-cumulative network representations, meaning that links represent only the transactions happening on that specific week, similarly to what is done in \cite{makarov2021blockchain} at monthly aggregation. This allows us to obtain a more representative structure for the state of the economy, avoiding considering in the same way connections that may have happened years apart from each other.

\section*{Acknowledgments}

C.C. acknowledges support from the Swiss National Science Foundation grant \#200021\_182659.
N.V. acknowledges support from the IOTA foundation.

\bibliography{biblio.bib}

\newpage
\appendix

\makeatletter 
\renewcommand{\thefigure}{S\@arabic\c@figure}
\makeatother
\makeatletter 
\renewcommand{\thetable}{S\@arabic\c@table}
\makeatother

\setcounter{figure}{0}
\setcounter{table}{0}

\section*{Supplementary Information: Robustness to address clustering}

In this document we present additional results that complement the main text, changing the address clustering algorithm to an alternative strategy that tends to aggregate less. These results provide a robustness check against the relatively arbitrary choice of heuristic algorithms that are used to identify entities, and in particular to the potential critical point that our results in the main text are tainted by excessive aggregation of addresses under the Combined Heuristics (CH) strategy.

For this reason we choose to present here the results adopting only the ``Multiple Inputs" (MI) heuristic described in the main text, which follows the logic that if multiple addresses appear as inputs to the same transaction, they are likely to belong to the same entity that had to sign them all to create the transaction.

In summary, our results are mostly qualitatively unchanged when changing the address clustering strategy. Here we highlight the key similarities and differences by referencing the tables and results from the main text in comparison to the ones presented in this Supplementary Information.

\begin{figure}[t]
 \centering
 \includegraphics[width=.95\linewidth]{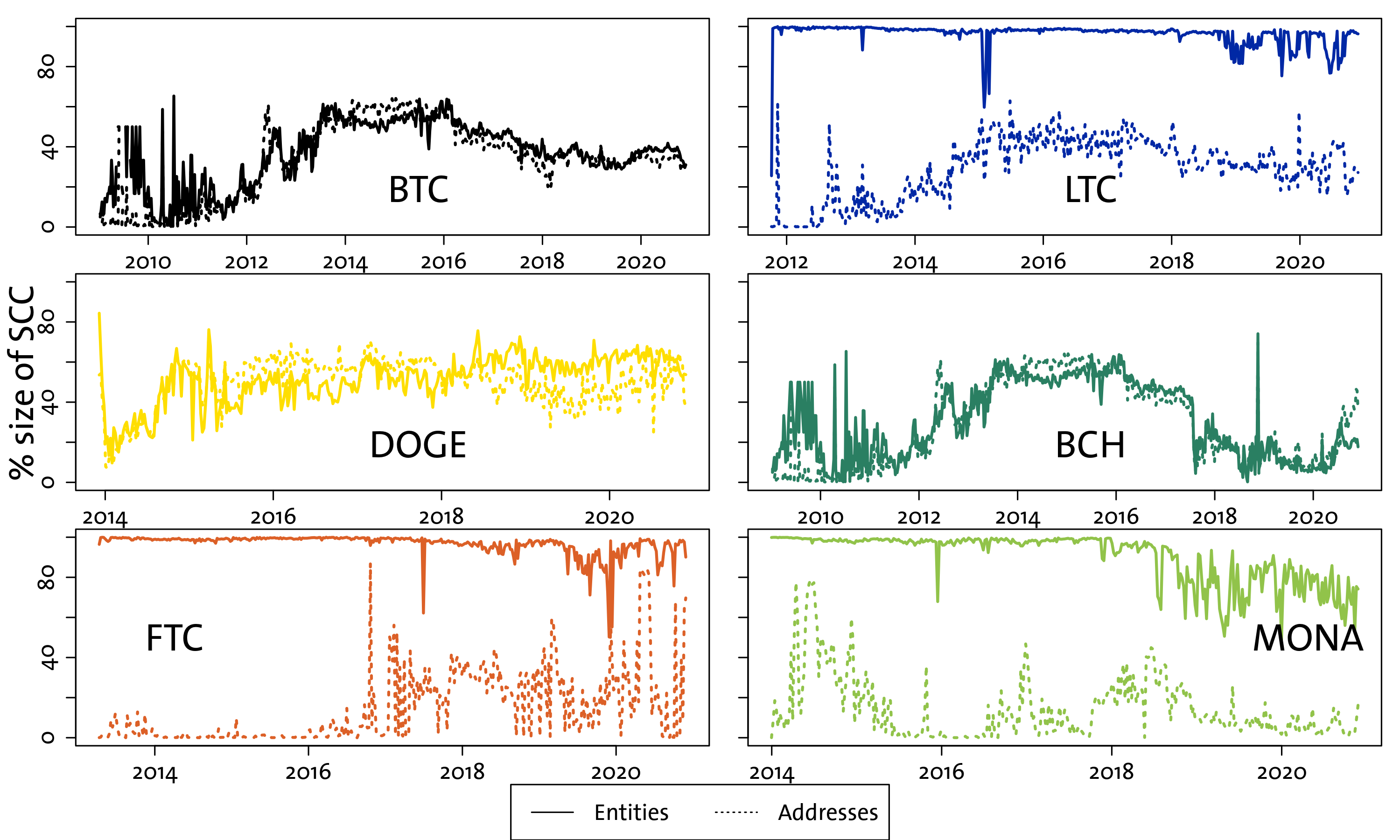}
 \caption{Size of the largest Strongly Connected Component in UTXO-based cryptocurrencies.}
 \label{fig:siscc}
\end{figure}

\begin{figure}[t]
    \centering
    \includegraphics[width=\textwidth]{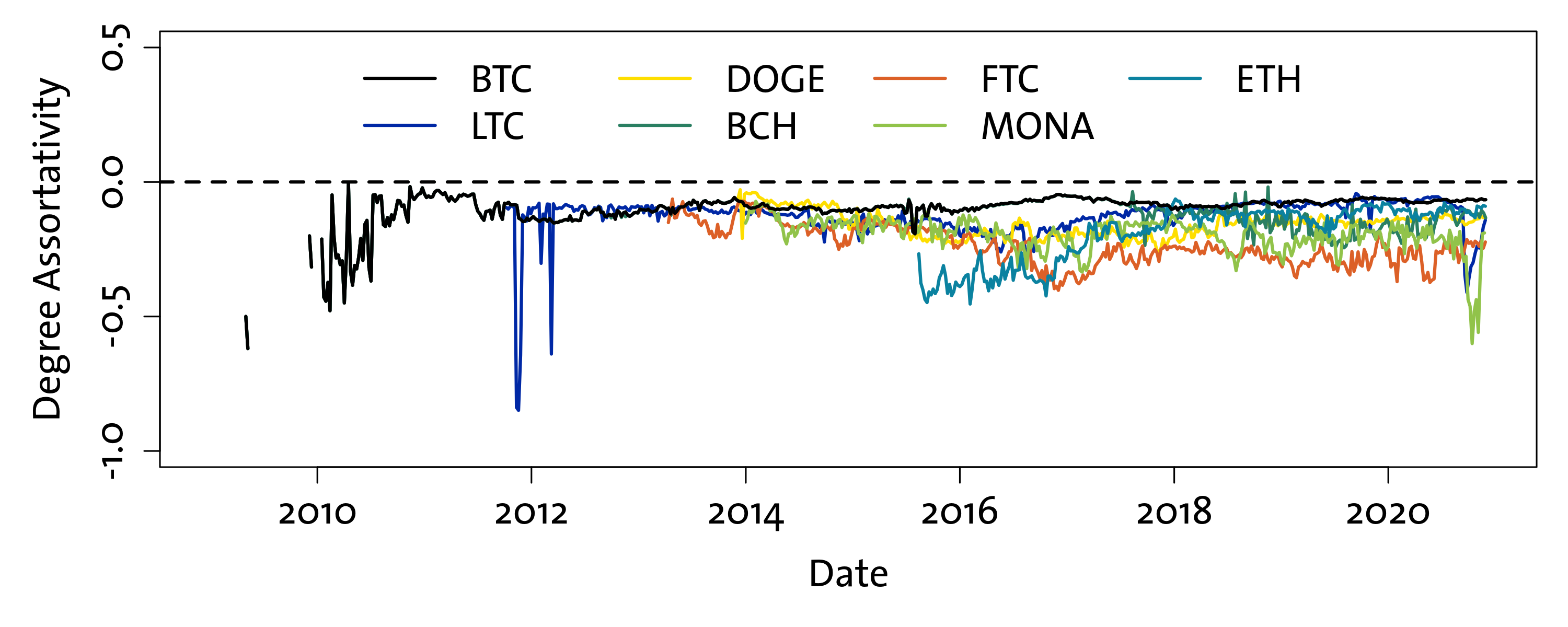}
    \caption{Assortativity coefficient for weekly transaction networks.}
    \label{fig:siassor}
\end{figure}

\begin{figure}[t]
    \centering
    \includegraphics[width=\textwidth]{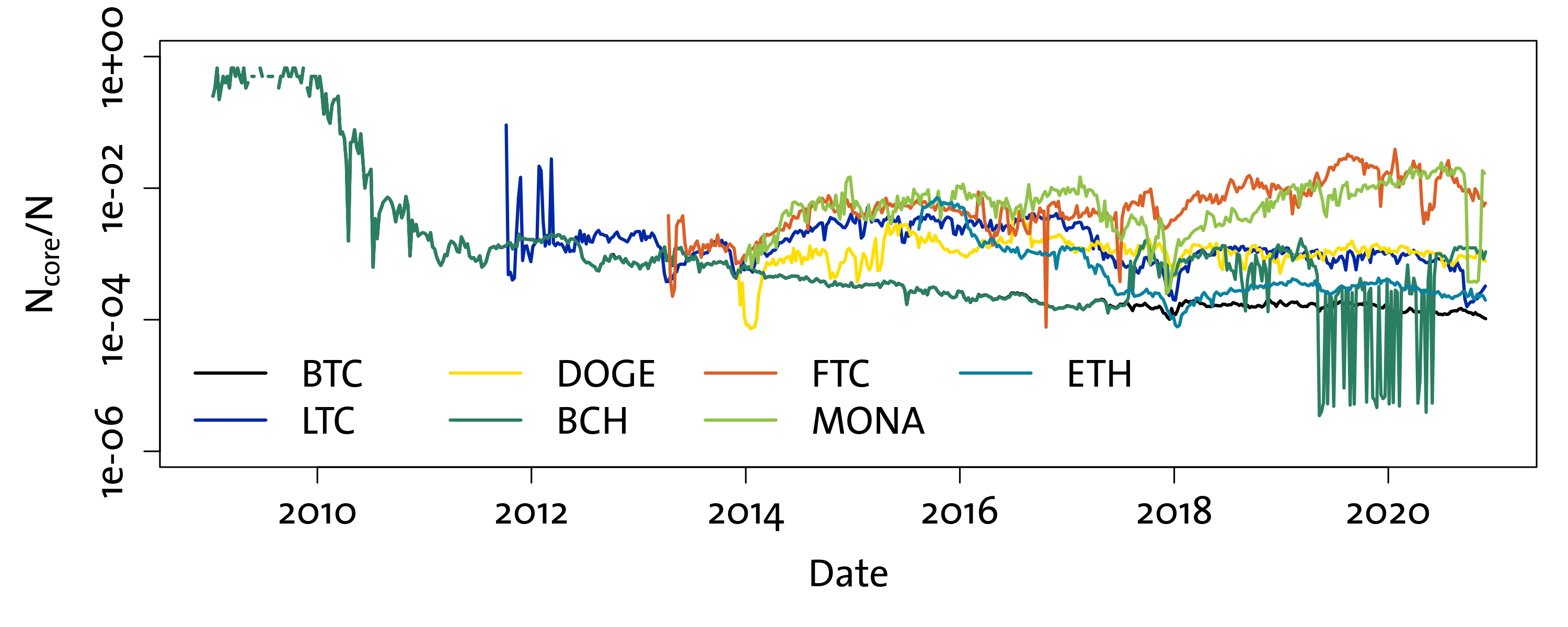}
    \caption{Fraction of nodes in the network core as a function of time.}
    \label{fig:sicoreperiph}
\end{figure}

\begin{figure}[t]
    \centering
    \includegraphics[width=\textwidth]{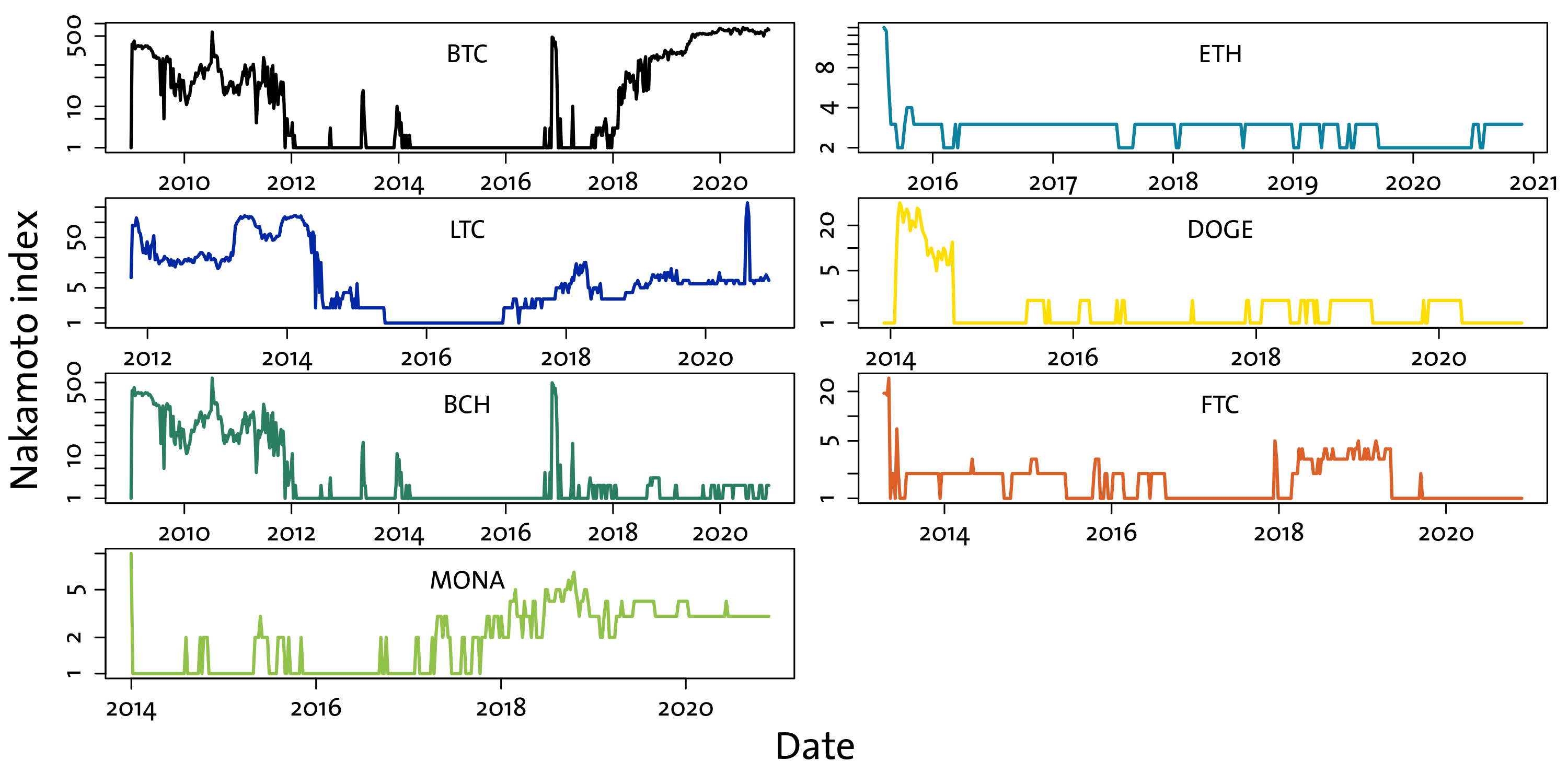}
    \caption{Nakamoto index estimated weekly for the analysed cryptocurrencies.}
    \label{fig:sinaka}
\end{figure}

\begin{itemize}
    \item in Figure \ref{fig:siscc} we show the comparison between the size of the largest Strongly Connected Component of the Addresses network and the Entities network under the MI heuristic. We see that it is qualitatively similar to Figure 2 of the main text for all coins with the notable exceptions of Bitcoin and Bitcoin Cash. Here the SCC of the MI Entities network does not span the entire set of nodes, whereas the CH Entities did;
    \item in Table \ref{tab:sidegexp} we report the equivalent of Table 1 in the main text for the MI Entities networks (except for ETH which is the same). We find that the fitted Pareto exponents are typically smaller and the fitted Pareto distributions are closer to the empirical ones according to the Kolmogorov-Smirnov distance, while the fitted binomials are typically worse fits, suggesting that degree distributions are even more fat-tailed in the case of MI Entities. Figure \ref{fig:siassor} shows the degree assortativity coefficient which we find to be slightly closer to $0$ than what we see for the CH Entities, although it still is always negative;
    \item The core size shown in Figure \ref{fig:sicoreperiph} is remarkably similar to the one shown in Figure 4 of the main text;
    \item The Nakamoto index, which in the main text is shown for Entities identified through the MI heuristic, is much smaller in the case of CH Entities as shown in Figure \ref{fig:sinaka}. We chose to report the ``safer" figure in the main text as the CH Entities might be subject to over-aggregation, which in this particular case would depict a more troublesome situation than the already worrying picture of Figure 5 in the main text.
    
\end{itemize}

\clearpage

\begin{table}[t]
\centering
\begin{tabular}{r|r|rrrrrrr}
  \hline
 & & BTC & LTC & DOGE & BCH & FTC & MONA & ETH \\ 
  \hline
\multirow{5}{*}{$\alpha$} & Minimum & 1.64 & 1.31 & 1.64 & 1.58 & 1.52 & 1.66 & 1.50 \\ 
  & Maximum & 4.93 & 3.66 & 3.36 & 4.93 & 5.67 & 3.97 & 3.42 \\ 
  & 1. Quartile & 2.04 & 1.87 & 1.86 & 1.96 & 1.93 & 1.97 & 1.68 \\ 
  & 3. Quartile & 2.46 & 2.52 & 2.33 & 2.43 & 2.76 & 2.21 & 1.81 \\ 
  %Mean & 2.71 & 2.77 & 2.78 & 2.65 & 2.74 & 2.85 & 1.82 \\ 
  & Median & 2.25 & 2.07 & 1.97 & 2.19 & 2.26 & 2.09 & 1.76 \\ 
   \hline
 $D_{Par}$ & Median & 0.03 & 0.05 & 0.04 & 0.03 & 0.07 & 0.06 & 0.03 \\ 
 $D_{Bin}$ & Median & 0.35 & 0.41 & 0.49 & 0.37 & 0.44 & 0.41 & 0.43 \\ 
\end{tabular}
\caption{Summary statistics on exponents of power-law fit of degree distribution on the weekly transaction networks.}\label{tab:sidegexp}
\end{table}

\end{document}